\documentclass[final,5p,times,twocolumn]{elsarticle}

\usepackage{graphics}

\usepackage{amssymb}

\journal{NIM B}

\begin{document}

\begin{frontmatter}



\title{Improved modelling of helium and tritium production for spallation targets}

\author[label1]{S. Leray}
\author[label1]{A. Boudard}
\author[label2]{J. Cugnon}
\author[label1]{J.C. David}
\author[label3]{A. Keli\'c-Heil}
\author[label2]{D. Mancusi}
\author[label3]{M.V. Ricciardi}
\address[label1]{CEA/Saclay, Irfu/SPhN, 91191 Gif-sur-Yvette, Cedex, France}
\address[label2]{University of Li\`ege, AGO Department, allée du 6 ao\^ut 17, B\^at. B5, B-4000 Liège 1, Belgium}
\address[label3]{GSI, Planckstrasse 1, D-64291 Darmstadt, Germany}



\begin{abstract}
Reliable predictions of light charged particle production in spallation reactions are important to correctly assess gas production in spallation targets. In particular, the helium production yield is important for assessing damage in the window separating the accelerator vacuum from a spallation target, and tritium is a major contributor to the target radioactivity. Up to now, the models available in the MCNPX transport code, including the widely used default option Bertini-Dresner and the INCL4.2-ABLA combination of models, were not able to correctly predict light charged particle yields. The work done recently on both the intranuclear cascade model INCL4, in which cluster emission through a coalescence process has been introduced, and on the de-excitation model ABLA allows correcting these deficiencies. This paper shows that the coalescence emission plays an important role in the tritium and $^3He$ production and that the combination of the newly developed versions of the codes, INCL4.5-ABLA07, now lead to good predictions of both helium and tritium cross sections over a wide incident energy range. Comparisons with other available models are also presented.

\end{abstract}


\end{frontmatter}


\section{Introduction}
\label{Intro}
Spallation reactions induced by high-intensity proton beams in a heavy-metal target are used to produce intense neutron fluxes for various applications such as spallation neutron sources, accelerator-driven sub-critical reactors, or radioactive ion beam facilities. These reactions produce large quantities of light charged particles (hydrogen and helium isotopes) which are a concern in spallation target design. Indeed, build up of gases (in particular helium) can lead to swelling and embrittlement of the window separating the accelerator vacuum and the spallation target and of other structural materials; tritium, a twelve-year half-life beta emitter, is a concern for radioprotection, especially in the case of liquid targets from which it can escape easily. It is therefore important that nuclear physics models implemented into the high-energy transport codes that are used by spallation target designers be able to reliably predict light charged particle production yields.

Up to now, light composite particle production was poorly predicted by nuclear models (generally a combination of an intranuclear cascade model followed by a de-excitation model) implemented into high-energy transport codes~\cite{RAP06, PIE06, LER06, BRO05}. In Ref.~\cite{RAP06} for instance, the cross sections for tritium production on iron and lead, as a function of the incident proton energy, were compared with MCNPX~\cite{MCN05} calculations using three different models: Bertini-Dresner~\cite{BER63, DRE62}, which is the default option, CEM2k from Ref.~\cite{MAS04} and Isabel~\cite{YAR79}-Dresner. It was found that these models were not able to account for tritium production, except Isabel-Dresner but only for lead. Discrepancies up to a factor of five were observed. Also, the dependence of the cross section with the incident energy is often not properly reproduced. In Ref.~\cite{PIE06}, experimental data from ~\cite{HIL01, HER06} on helium and hydrogen production cross sections as a function of the target charge, at different energies, were compared with FLUKA~\cite{FLU02}. A systematic underprediction by the model, more important for helium and heavy targets, was reported. In~\cite{BRO05}, a systematic comparison of many different models with alpha production cross section from tantalum, tungsten and gold was performed, which also pointed out large discrepancies between the calculation results and the experimental data.

During the last ten years, in Europe, an important effort has been devoted to the collection of high quality experimental data and, simultaneously, to the development of improved spallation models~\cite{HIN05}. As a result of this work, the combination of the intranuclear cascade model, INCL4.2~\cite{BOU02}, and de-excitation model, ABLA~\cite{JUN98}, has been developed, tested against a large set of experimental data and shown to give globally better predictions than models used by default (Bertini-Dresner~\cite{BER63, DRE62}) in high-energy transport codes such as MCNPX~\cite{MCN05}. INCL4.2-ABLA is now available in MCNPX and GEANT4~\cite{AGO03}. With these models, the situation regarding the predictions of neutron emission, heavy evaporation and fission residues production could be considered as having been largely improved. However, important deficiencies were still remaining when light charged particles are concerned. For helium out of iron, a systematic underprediction by a factor 2 to 3 by INCL4-ABLA was reported in Ref.~\cite{LER06}. Since in ABLA only nucleons and alphas can be evaporated, INCL4-ABLA does not produce any tritium. This is obviously a major shortcoming since, for instance in calculations of radioactivity of the MEGAPIE target~\cite{LEM07}, tritium has been identified as a major contributor around 10 years after irradiation when using models as Bertini-Dresner or CEM.

New versions of INCL4 and ABLA (INCL4.5~\cite{BOU08, CUG09} and ABLA07~\cite{KEL08, RIC09} respectively) have recently been released, which, among other improvements, were developed to remedy this situation: cluster emission, already introduced in the intranuclear cascade model in Ref~\cite{BOU04} to account for the high-energy tail observed in the experiments, has been refined; evaporation of deuterium, tritium and $^3 He$ is now taken into account in ABLA together with the possibility of emitting intermediate mass fragments, and barriers for light charged particle emission have been established on better physics basis. In this paper, we present results concerning improved predictions of helium and tritium obtained with these new versions. The work on the models has been done in the framework of the NUDATRA domain of the EUROTRANS European FP6 project~\cite{NUD04}, whose objective is to provide improved simulation tools for the design of ADS transmuters.

\section{Cluster emission in INCL4}
\label{Sec2}
\begin{figure}[hbt]
\begin{center}
\includegraphics[width=6.3cm]{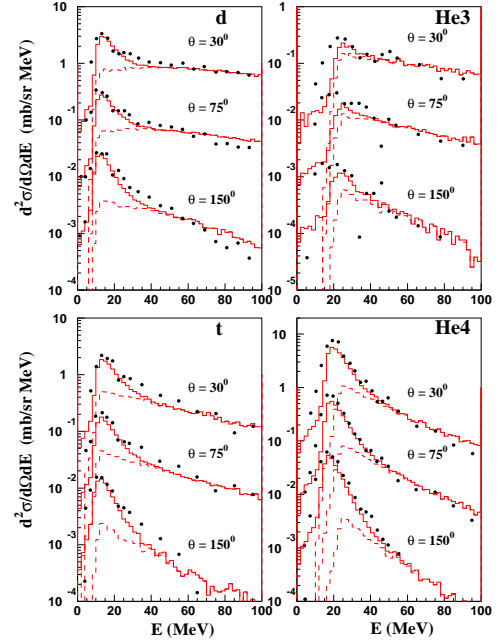}
\end{center}
\caption{\label{Fig1}
Light composite particle double-differential cross sections, in the p+Ta reaction at 1.2 GeV, measured by ~\cite{HER06} and compared with the results of INCL4.5-ABLA07. The cross sections are mutliplied by $10^0$, $10^-1$ and $10^2$, successively, starting from the smallest angle. The contribution due to the coalescence process in INCL4.5 is given by the dashed line.}
\end{figure}

In light composite particle energy spectra, two different components are generally observed, as shown by the experimental data from Herbach et al.~\cite{HER06} plotted in Fig.~\ref{Fig1}: a low-energy isotropic peak coming from the evaporation stage, and a high-energy forward-peaked tail. An attempt to separate the two components was done by the authors of Ref.~\cite{HER06}. They extracted the relative contribution of this high-energy component, called pre-equilibrium in their paper, to the total yields of the different light composite particles, as a function of the target charge, at 1.2 GeV. This showed that, while $^4 He$ is produced predominantly by evaporation, for the other light composite particles the so-called pre-equilibrium contribution is far from being negligible, reaching even 60$\%$ for $^3 He$ on high-Z targets.

The level of this contribution in the production cross sections of tritium and $^3 He$ shows how important it is to be able to account for the high-energy tail with the models. This is the reason why a coalescence mechanism leading to light composite particle emission has been implemented into INCL4, first in Boudard et al.~\cite{BOU04}, and then modified in order to get a better ratio between the different species of clusters~\cite{BOU08, CUG09}. The cluster emission mechanism is based on surface coalescence in phase space, i.e. on the assumption that a cascade nucleon ready to escape at the nuclear surface can coalesce with other nucleons close enough in phase space and form a cluster. The parameters of the model include the volume of the phase space cell in which nucleons should be to form a cluster and the distance from the surface at which the clusters are built. All possible clusters up to a given mass number are formed and the priority is given to the one with the lowest excitation energy per nucleon. The selected cluster is emitted only if it succeeds to go through the Coulomb barrier. In INCL4.5, used in this paper, clusters up to mass 8 are considered.

The other modifications compared to the original INCL4 model~\cite{BOU04} involves the introduction of a potential for pions and of an isospin and energy dependent nuclear potential~\cite{BOU08}. Despite the fact that intranuclear cascade models are based on assumptions that are in principle not valid below 100-150 MeV, INC models are nevertheless used in transport codes below these energies, for instance when evaluated data files for the considered nuclei do not exist. Therefore, it is important that the models give reasonable results also at low energies. In the case of INCL4.5, in order to get better results at low energies, in particular concerning the total reaction cross section, special attention has been paid to the treatment of the first nucleon-nucleon collision and Coulomb distortion in the entrance channel has been introduced~\cite{CUG09}.

\begin{figure}[h]
\begin{center}
\includegraphics[width=6.3cm]{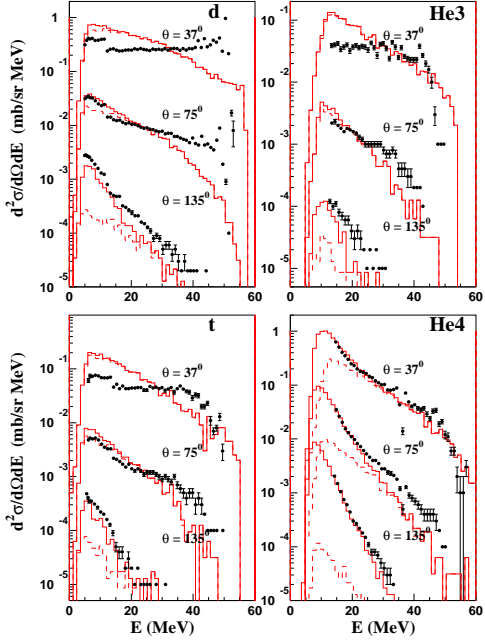}
\end{center}
\caption{\label{Fig2}
Same as Fig.~\ref{Fig1} but for the p+Fe reaction at 62 MeV, measured by~\cite{BER73}.}
\end{figure}

In Fig.~\ref{Fig1}, results on light composite particle double-differential cross sections obtained with INCL4.5 coupled to the de-excitation model ABLA07~\cite{KEL08} are compared with the experimental data from~\cite{HER06} measured at 1.2 GeV in the p+Ta reaction. The dashed curve shows the contribution coming from the coalescence process in INCL4.5. It can be seen that the high energy tail is very well reproduced by this process for all species of light charged particles, although a slight overprediction of $^3 He$ and $^4 He$ at the highest particle energies can be observed. The  angular dependence is also reasonably reproduced. It can be noticed, as already realized by the authors of Ref.~\cite{HER06}, that the cascade contribution to the total cross section (dashed curve) is very important in the case of tritium and $^3 He$ contrary to the case of $^4 He$. A similar agreement, not shown here, is obtained when comparing the model with p+Au data from Ref.~\cite{LET00} at 2.5 GeV (this and further results can be found in the benchmark of spallation models recently organized by IAEA~\cite{IAEA09}). 

Fig.~\ref{Fig2} gives an example of the results obtained at low incident energies, here 62 MeV for the system p+Fe compared to the data from~\cite{BER73}. The agreement is not as good as at high incident energies, in particular at forward angles where the slopes of the calculated energy spectra are too steep,  but still satisfactory. At this incident energy, the cascade coalescence contribution is clearly dominant except at backward angles for d, t, $^3 He$ and for $^4 He$ at all angles. It should be stressed that the good agreement achieved for different target masses on a wide range of incident energies is obtained with the same set of parameters in the coalescence model, chosen once for all.

\section{Recent improvements of ABLA}
\label{Sec3}
Simultaneously to the work on INCL4, the authors of ABLA have produced a new improved version of the model, ABLA07~\cite{KEL08}. This version now allows evaporation of all the types of light charged particles from p to $^4 He$ but also of intermediate mass fragments. It uses improved parameterizations of inverse reaction cross sections and Coulomb barriers in order to better reproduce experimental particle energy spectra. A simultaneous break-up (multifragmentation) mechanism has been added for systems overcoming a certain limiting temperature~\cite{NAT02}. The fission part has also been modified. More details as well as examples of the improved agreement with experimental data can be found in Ref.~\cite{RIC09}. 

It can be seen in Fig.~\ref{Fig1} that the low energy part of the deuterium and tritium spectra, corresponding to the evaporation peak, is very well reproduced by ABLA07. For $^3 He$, the situation is less good since the production yield is a little too small and the evaporation peak is not broad enough, especially on the low-energy side. $^4 He$ is intermediate between tritium and $^3 He$. Since INCL4.5-ABLA07 also agrees very well with the neutron evaporation spectra measured on Pb at 1.2 GeV (results not shown here but which can be found in the IAEA benchmark~\cite{IAEA09}), we are rather confident that excitation energy distribution at the end of the cascade stage is correct. Therefore, the slight discrepancy observed for helium isotopes is more likely an indication that the barriers and/or tunneling through the barrier in case of the helium emission in ABLA07 still have to be improved.

\section{Total tritium production yields}
\label{Sec4}

\begin{figure}[h!]
\begin{center}
\includegraphics[width=6.3cm,angle=-90]{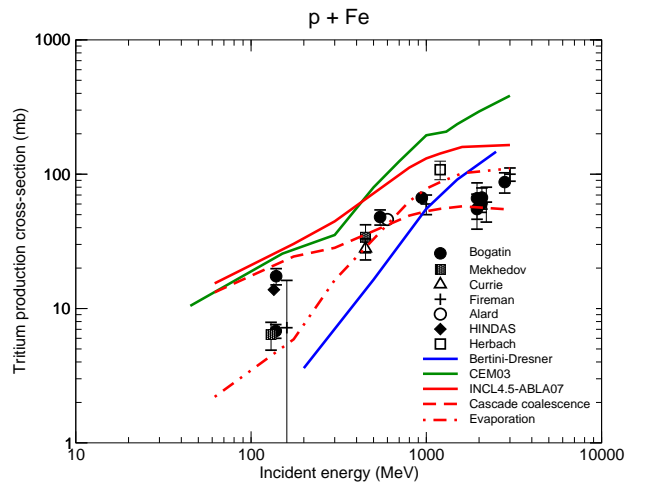}
\end{center}
\caption{\label{Fig3}
Tritium production cross sections in iron calculated with the new versions INCL4.5-ABLA07 (solid red line) compared to data measured by different groups (Refs.~\cite{BOG76, MEK70, CUR56, FIR57, ALA75, HIN05, HER06}) and to calculations with Bertini-Dresner (blue) and CEM03 (green) models. The dashed, resp. dashed-dotted, lines show the respective contributions from the cascade coalescence and evaporation processes.}
\end{figure}

\begin{figure}[h!]
\begin{center}
\vspace*{0.3 cm}
\includegraphics[width=6.3cm,angle=-90]{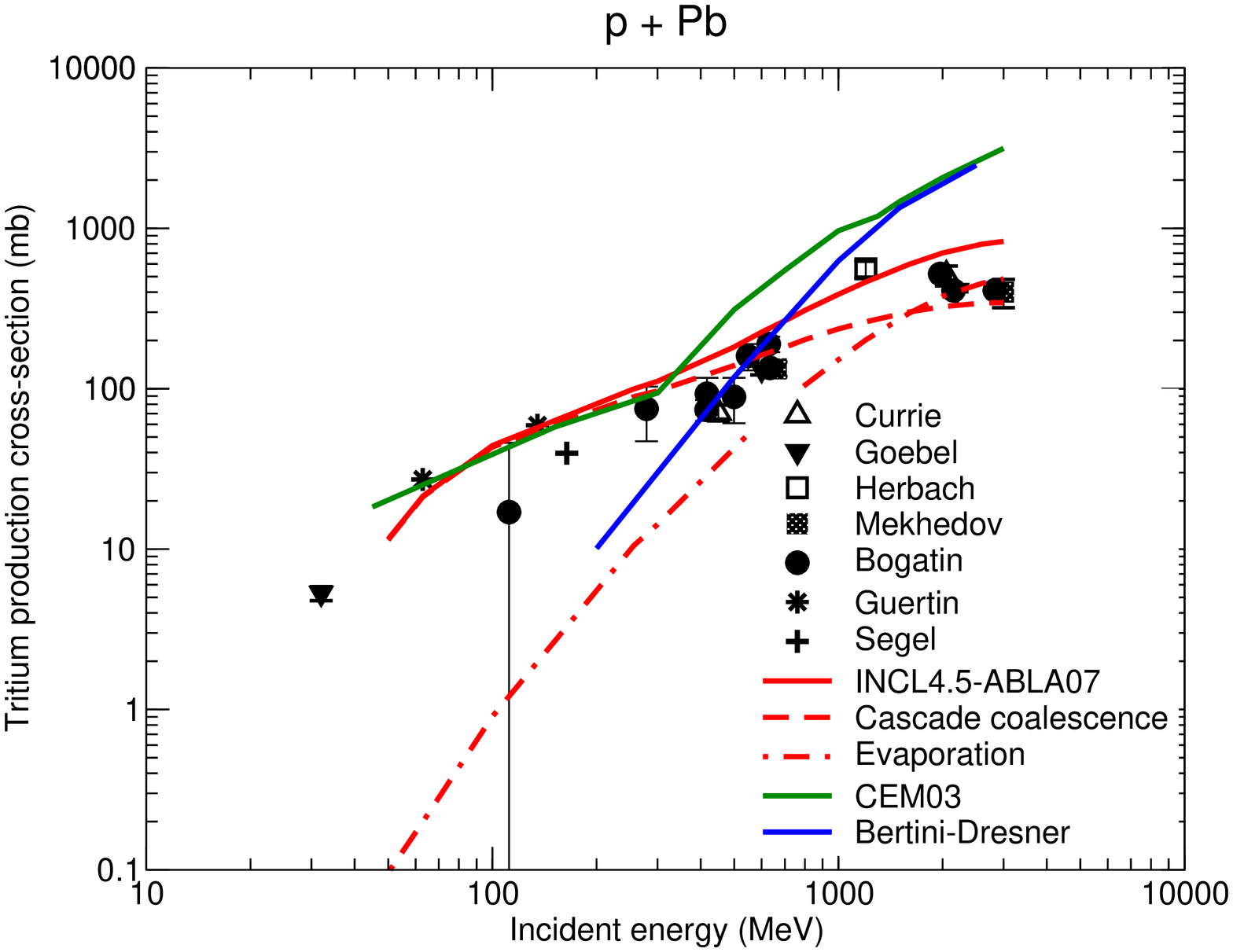}
\end{center}
\caption{\label{Fig4}
Same as Fig.~\ref{Fig3} but for lead. Data from Refs.~\cite{BOG76, MEK70, CUR56, GOE64, HIN05, GUE05, HER06}.}
\end{figure}

Figs.~\ref{Fig3} and~\ref{Fig4} present the results obtained with INCL4.5-ABLA07 (solid red line) for the total production yields of tritium on iron and lead targets, respectively, as a function of the incident proton energy, compared to the available experimental data~\cite{BOG76, MEK70, CUR56, FIR57, ALA75, GOE64, GUE05, HER06, HIN05}. Calculations with other models implemented in MCNPX, namely Bertini-Dresner (blue line), which is the default option, and the new version of the CEM model (green line) from S. Mashnik, CEM03~\cite{MAS06}. For both iron and lead, INCL4.5-ABLA07 gives a good agreement with the data all along the energy range. This combination of models is definitely better than Bertini-Dresner which has a totally wrong energy dependance and even than CEM03 which largely overestimates tritium production at high incident energies.  

The respective contributions of the cascade coalescence (dashed line) and evaporation processes (dashed-dotted line) in the tritium production cross sections are also shown in the figures. It can be seen that, at low energies, tritium is produced nearly exclusively by the coalescence process, evaporation becoming dominant only above 600 MeV for Fe and 1.7 GeV for Pb. This illustrates the importance of having models able to account for the non-evaporative component of the spectra and explains why models such as Bertini-Dresner, or ISABEL-Dresner not discussed in this paper, cannot give good predictions, especially at low incident energies. In CEM03, high-energy clusters are produced either in a coalescence process or in the pre-equilibrium stage.

\section{Helium production yields}
\label{Sec5}

\begin{figure}[h!]
\begin{center}
\vspace*{0.3 cm}
\includegraphics[width=6.3cm,angle=-90]{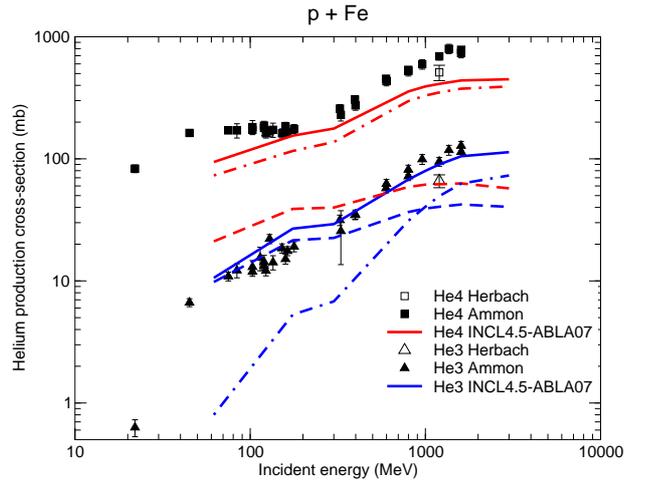}
\end{center}
\caption{\label{Fig5}
$^3 He$ (blue) and $^4 He$ (red) production cross sections in iron calculated with INCL4.5-ABLA07 compared to experimental data from~\cite{HER06, AMM08}. The dashed, resp. dashed-dotted, curves give the contribution from the cascade coalescence, resp. evaporation emission.}
\end{figure}

Fig.~\ref{Fig5} shows the result for the $^3 He$ and $^4 He$ production cross section in iron as a function of the proton incident energy calculated with INCL4.5-ABLA07, with the respective contributions of cascade coalescence and evaporation, compared to the experimental data from~\cite{HER06, AMM08}. Like tritium, $^3 He$ is predominantly originating from the cascade coalescence process up to about 1 GeV. On the other hand, $^4 He$ is produced essentially by evaporation, the cascade contribution representing at most 20$\%$ at the lowest energies. This illustrates the difference in the two physical mechanisms: in the evaporation process, the emission of tightly bound particles (alphas) is favored compared to more loosely bound ones (d, t, $^3 He$), while the coalescence is mainly governed by the number of nucleons in a cluster (there are more clusters of mass 2 than 3, 3 than 4), the binding energy of the clusters playing only a minor role.   

\begin{figure}[h!]
\begin{center}
\vspace*{0.3 cm}
\includegraphics[width=6.3cm,angle=-90]{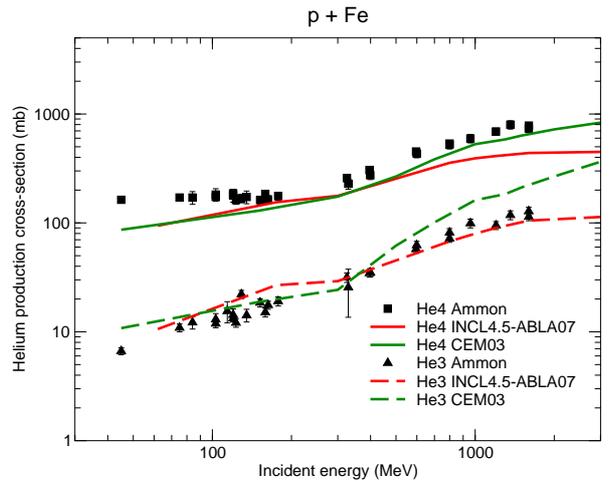}
\end{center}
\caption{\label{Fig6}
$^3 He$ (dashed) and $^4 He$ (solid) production cross sections in iron calculated with INCL4.5-ABLA07 (red) compared to data measured by different groups (Refs.~\cite{HER06, AMM08}) and to calculations with CEM03 (green).}
\end{figure}

The combination of the two models reproduce very well the $^3 He$ cross section on the full energy range while it underpredicts $^4 He$ a little bit. Since $^4 He$ is predominantly produced by evaporation, this underestimation could be related with the fact, mentioned in section~\ref{Sec3}, that ABLA07 gives a too narrow evaporation peak for helium isotopes. However, this observation was made for Ta and the situation could be different for a much lighter nucleus like Fe. Unfortunately, the double-differential cross sections on Fe measured by the NESSI group are not available. In  Fig.~\ref{Fig6}, the results given by the CEM03 code are also compared to the experimental data and to our model. It can be seen that, at high incident energies, CEM03 overpredicts $^3 He$, as it was the case for tritium, but is slightly better than INCL4.5-ABLA07 for $^4 He$.  

\begin{figure}[h!]
\begin{center}
\vspace*{0.3 cm}
\includegraphics[width=6.3cm,angle=-90]{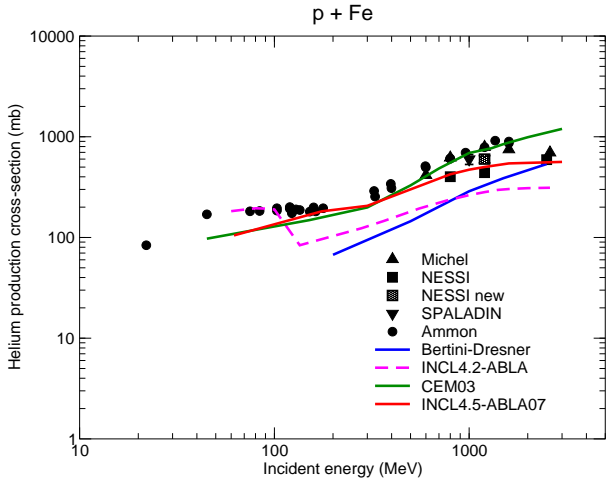}
\end{center}
\caption{\label{Fig7}
Helium production cross sections in iron calculated with INCL4.5-ABLA07 (red) compared to data measured by ~\cite{HER06, AMM08} and to calculations with INCL4.2-ABLA (dashed pink), CEM03 (green) and Bertini-Dresner (blue) models.}
\end{figure}

Since what matters for material damage assessment is the total helium production yield, the helium production cross section is shown in Fig.~\ref{Fig7} for different models compared to the available experimental data. The first remark is that the new version, INCL4.5-ABLA07, represents a clear improvement compared to the previous one, INCL4.2-ABLA, and to Bertini-Dresner, which, being the default option of MCNPX, is used by most of the users of the code. In INCL4.2-ABLA the step observed around 100 MeV was due to a forced absorption process added at low incident energies in order to obtain the right total reaction cross section. In the new model this is no longer necessary since it naturally gives the correct total reaction cross section, as mentioned in section~\ref{Sec2}. As already observed, CEM03 predicts somewhat larger cross sections above 500 MeV than our model. In view of the discrepancy between the different sets of experimental data, it is difficult to conclude which model is the best.

\begin{figure}[h!]
\begin{center}
\vspace*{0.3 cm}
\includegraphics[width=6.3cm,angle=-90]{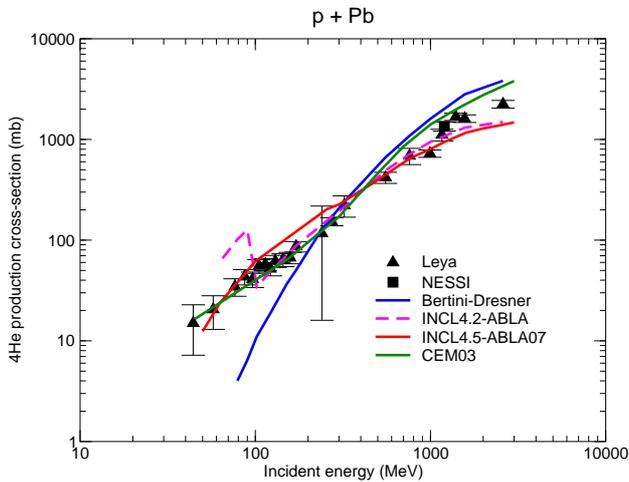}
\end{center}
\caption{\label{Fig8}
$^4 He$ production cross sections in lead calculated with INCL4.5-ABLA07 (red) compared to data measured by ~\cite{ENK99, HER06, LEY05, LEG08} and to calculations with INCL4-ABLA (dashed pink), CEM03 (green) and Bertini-Dresner (blue) models.}
\end{figure}

Regarding $^4 He$ production on lead, Fig.~\ref{Fig8} shows the comparison of our model with the experimental data from~\cite{ENK99, HER06, LEY05, LEG08} and other models. As for iron, INCL4.5-ABLA07 and CEM03 give the best agreement with the experimental data, CEM03 producing a little more $^4 He$ than our model at high incident energies. Bertini-Dresner has a much too steep dependence with incident energy, leading to a severe underestimation of helium at low energies. Compared to the INCL4.2-ABLA version, our new model gives similar results except at low incident energies where the step around 100 MeV has been suppressed. 

\begin{figure}[h!]
\begin{center}
\vspace*{0.3 cm}
\includegraphics[width=6.3cm,angle=-90]{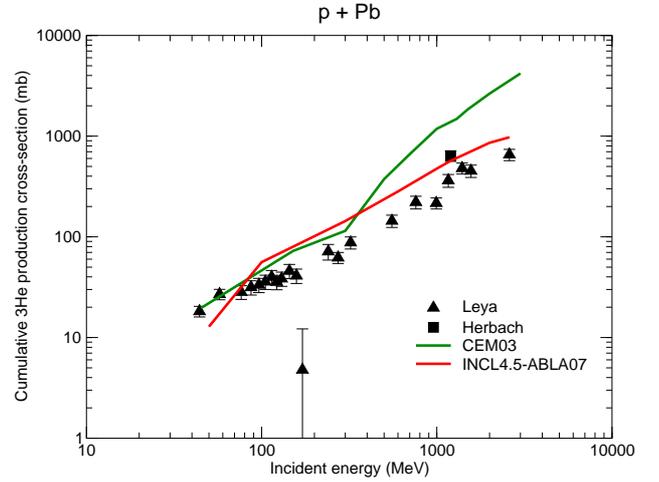}
\end{center}
\caption{\label{Fig9}
Cumulative $^3 He$ production cross sections in lead calculated with INCL4.5-ABLA07 (red) compared to data measured by~\cite{ENK99, HER06, LEY05, LEG08} and to calculations with CEM03 (green).}
\end{figure}

There do not exist many data on the production of direct $^3 He$ from lead. Therefore, we have compared the models to the measurement by Leya et al.~\cite{LEY05} of cumulative yields of $^3 He$, i.e. direct $^3 He$ plus $^3 He$ from the decay of tritium, as a function of incident energies. The results are shown in Fig.~\ref{Fig9} in which the calculated cross sections by INCL4.5-ABLA07 and CEM03 of $^3 He$ and tritium have been added up to be compared to the experimental data. The same was done for the Herbach data point displayed in the figure. The same conclusions as for tritium production in Fig.~\ref{Fig4} can be drawn, i.e. that our model gives a better agreement to the data than CEM03. This actually is not surprising since tritium represents around 80$\%$ of the cumulative yield. This observation is rather an indication that the different sets of experimental data are consistent. 

\section{Conclusion}

In this paper, we have compared the predictions of the newly developed versions of the INCL4 and ABLA models, INCL4.5-ABLA07, to the helium and tritium production double-differential cross sections and excitation functions found in the literature. It has been shown that the combination of the new versions represent a definite improvement compared to the former models, which did not evaporate tritium and $^3 He$ and were unable to account for the high energy tail of the particle spectra. In particular, it was found that the coalescence process, added in the cascade model to describe the emission of high energy clusters, is very important to get correct production cross sections of tritium and $^3 He$ since it represents the major part of the cross section up to rather high energies. 

Comparisons with the other models available in MCNPX, Bertini-Dresner and CEM03, have also been shown, which definitely rules out the first one although it is the most often, blindly, used (default) option.  CEM03 was found to be less good than our model on tritium and $^3 He$ production, especially at incident energies above 300-400 MeV. For $^4 He$ or total helium cross sections, it is difficult to conclude which of the two models is the best in view of the discrepancies between the different sets of data. 

The INCL4.5-ABLA07 combination of models will be soon available in MCNPX~\cite{DAV09}. In view of its validation over the wide energy range presented in this paper, this should allow reliable simulations of, for instance, tritium build up in liquid metal spallation targets or gas production in material structures. Also, the good prediction of the high-energy tail of the emitted light-charged particles, should guaranty a reasonable assessment of the secondary reactions potentially induced by these particles.

{\bf
\noindent
Acknowledgments}

This work was partly supported by the FP6 Euratom project EUROTRANS/NUDATRA, EC contract number FI6W-CT-2004-516520.

\end{document}